\documentstyle[twoside,fleqn,espcrc2]{article}

\input epsf

\newcommand{\AmS}{{\protect\the\textfont2
  A\kern-.1667em\lower.5ex\hbox{M}\kern-.125emS}}

\title{ Monopole Condensation and Polyakov Loop 
in Finite-Temperature Pure QCD }

\author{Shinji Ejiri 
\address{
Department of Physics, Kanazawa University, Kanazawa 920-11, Japan}
}

\begin{document}

\begin{abstract}
We study the relation between the abelian monopole condensation and 
the deconfinement phase transition of the finite-temperature pure QCD.
The expectation value of the monopole contribution to the Polyakov loop 
becomes zero when a long monopole loop is distributed uniformly 
in the configuration of the confinement phase. 
On the other hand, it becomes non-zero when the long monopole 
loop disappears in the deconfinement phase. 
We also discuss the relation  between the monopole behaviors and 
the usual interpretation of the spontaneous breaking of Z(N)
symmetry in finite-temperature SU(N) QCD.
It is found that the boundary condition of the space direction is important 
to understand the Z(N) symmetry in terms of the monopoles.

\end{abstract}

\maketitle

\section{Introduction}

The purpose of this study is 
to confirm if the quark confinement mechanism in QCD can be understood 
as dual Meissner effect due to abelian monopole condensation. 
This idea is supported by 
recent results of Monte-Carlo simulations of abelian projected QCD. 
The remarkable results in finite-temperature QCD is the followings: 
(1) The configuration of the monopole currents changes dramatically 
at $ T_{c} $ when we adopt the maximally abelian gauge. 
(2) The monopole contribution to the Polyakov loop shows similar behavior 
to the original Polyakov loop and the residual photon part of the 
Polyakov loop is finite even in the confinement phase. 
It means that the monopole part controls the deconfinement transition 
\cite{suzu}.
(3) Last year, we found that the behavior of the Polyakov loop 
can be explained by the monopole dynamics, 
considering the 3-dimensional solid angle in SU(2) gauge theory 
\cite{ejiri}. 

Let us explain briefly the interpretation 
using the 3-dimensional solid angle.
First, updation of link variables is done as usual.
Then, we perform abelian projection \cite{thooft,kron} 
and extract the abelian gauge field.
\begin{eqnarray}
 U'_{\mu}(s) &=& C_{\mu}(s) \ u_{\mu}(s), \\
& & \hspace{-15mm} 
u_{\mu} = {\rm diag} ( e^{i \theta^{(1)}_{\mu}}, \cdots, 
e^{i \theta^{(N)}_{\mu}}),
\hspace{5mm} {\rm det} u_{\mu} =1 , \nonumber
\end{eqnarray}
where 
$ U'_{\mu} $ is the gauge fixed link field, 
$ u_{\mu} $ is the abelian link field, and  
$ \theta^{(i)}_{\mu} $ can be regarded as the abelian gauge field. 
We adopt the maximally abelian gauge. 
The monopole current $ k_{ \mu }(s) $ is defined as 
\begin{eqnarray}
k_{\mu}(s) \hspace{-2mm} &=& \hspace{-2mm} 
(1/4\pi)\epsilon_{\mu\alpha\beta\gamma}\partial_{\alpha}
\bar{\Theta}_{\beta\gamma}(s), \\
\Theta_{ \mu \nu }(s) \hspace{-3mm} &=& \hspace{-3mm}
\partial_{\mu} \theta_{\nu} (s)- \partial_{\nu} \theta_{\mu} (s) 
= \bar{\Theta}_{ \mu \nu }(s) + 2 \pi n_{\mu \nu}(s), \nonumber
\end{eqnarray}
where $ \bar{\Theta}_{\mu \nu} \in [-\pi, \pi) $ and 
$ n_{\mu \nu} $ is an integer,
following DeGrand-Toussaint \cite{degrand}. 
Here, $ ^{\ast} \hspace{-1mm} n_{\mu \nu} (s) = 
\frac{1}{2} \epsilon_{\mu \nu \rho \sigma} 
n_{\rho \sigma}(s) $ 
is the Dirac string: 
$ k_{\mu} = \partial_{\nu} \  ^{\ast} \hspace{-1mm} n_{\nu \mu} $.
The abelian Polyakov loop operator $ P_{a} $ and the contributions from 
the monopole $ P_{m} $ and the photon $ P_{p} $ are defined as follows 
\cite{suzu}: 
\begin{eqnarray}
P_{a} \hspace{-3mm} &=& \hspace{-3mm} \frac{1}{N} \hspace{-1mm} 
\sum_{i}\exp \{i \sum_{j=0}^{N_{t}-1} \theta^{(i)}_4 (s+j\hat4) \} 
= P_{p} \cdot P_{m}, \\
P_{p} \hspace{-3mm} &=& \hspace{-3mm} \frac{1}{N} \hspace{-1mm} 
\sum_{i} \hspace{-1mm} \exp\{-i \hspace{-1mm} 
\sum_{j,s'} \hspace{-1mm} 
D(s+j\hat4-s')\partial'_{\nu}\bar{\Theta}^{(i)}_{\nu 4}(s')\}, \\
P_{m} \hspace{-3mm} &=& \hspace{-3mm} \frac{1}{N} \hspace{-1mm} \sum_{i} 
\hspace{-1mm} \exp\{-2\pi i \hspace{-1mm} \sum_{j,s'} \hspace{-1mm} 
D(s+j\hat4-s')\partial'_{\nu}n^{(i)}_{\nu 4}(s')\}, \nonumber  
\vspace{-5mm} \\
\end{eqnarray}
where $D(s-s')$ is the lattice Coulomb propagator satisfying:
$ \partial'_{\mu} \partial_{\mu} D(s) = - \delta_{s,0} $. 

Here we restrict ourselves to the monopole part. 
If we integrate over the time direction, the monopole part of 
the Polyakov loop can be rewritten in the infinite-volume limit 
as follows: 
\begin{eqnarray} 
P_{m} (\vec{s}) = \exp(2 \pi i \frac{\Omega(\vec{s})}{4 \pi}), 
\end{eqnarray}
where $\Omega(\vec{s})$ is the oriented solid angle 
made by monopole loops looking from Polyakov loop ($\vec{s}$) 
in the 3-dimensional reduced space. 
The properties of Polyakov loop can be understood 
by the monopole dynamics qualitatively.

In the confinement phase ${ \rm ( T < T_{c} ) }$, 
long monopole loops distribute uniformly \cite{kita}. 
The $\Omega$ can be random from $0$ to $4 \pi $. 
The Polyakov loop at each point is random. 
The average of the Polyakov loop is zero.
On the other hand, 
in the deconfinement phase ${ \rm ( T > T_{c} ) }$, 
there is a space where no monopole exists. 
In such a space,  $ \Omega $ takes small value, 
and the local Polyakov loops are nearly one. 
Hence, the average of the Polyakov loop approaches one. 
This picture is confirmed by the measurement of 
the histogram of the solid angle using Monte-Carlo simulations. 
However, the effect of the boundary condition in the finite-volume 
was not considered in that explanation. 
Here we study the relation between the monopole dynamics 
and the Z(N) symmetry that controls the behavior of the 
Polyakov loop in SU(2) and SU(3) gauge theory. 
We found that the boundary condition in the finite volume analysis 
is very important to understand the Z(N) symmetry in terms of the monopole.

\section{Boundary Conditions}

When the periodic boundary condition is adopted,
the lattice Coulomb propagator can not be defined exactly. 
We use the following approximation for the Coulomb propagator: 
\begin{eqnarray}
 \partial_{\mu}' \partial_{\mu} D(s) 
= - \delta_{s,0} + \frac{1}{N_{\rm site}}, 
\end{eqnarray}
where $ N_{\rm site} $ is the number of sites. 
Notice that there exist many closed Dirac sheets (Bubbles of Dirac sheet). 
A bubble of Dirac sheet does not contribute 
to the Polyakov loop when the Coulomb propagator is exact. 
However, in the periodic boundary condition, 
the bubble of Dirac sheet gives the finite phase of 
the monopole part of the abelian Polyakov loop 
due to the incompleteness of the propagator: 
\begin{eqnarray}
P_{m} = \exp \{ 2 \pi i \frac{N_{b}}{N_{\rm space}} \}, 
\end{eqnarray}
where $ N_{\rm space} $ and $ N_{b} $ are the number of sites and 
the number of sites inside a bubble in 3-dimensional space respectively.
It is found that the case corresponding to the Z(N) phase, i.e., 
$ N_{b} / N_{\rm space} = k/N \ (k: {\rm integer}) $ alone 
keeps the action invariant and then is important.

Next, let us consider the anti-periodic boundary condition 
(C-periodic boundary condition). 
The lattice Coulomb propagator can be defined exactly 
by imposing the anti-periodic boundary condition for the space direction: 
\begin{eqnarray}
D(s+N_{s} \hat{i}) = -D(s), \hspace{2mm} 
\partial_{\mu}' \partial_{\mu} D(s) = - \delta_{s,0}. 
\end{eqnarray}
Then the gauge fields and the monopole currents must satisfy 
the anti-periodic boundary condition: 
\begin{eqnarray}
\theta_{\mu} (s+N_{s} \hat{i}) = - \theta_{\mu} (s), 
\ k_{\mu} (s+N_{s} \hat{i}) = - k_{\mu} (s) 
\end{eqnarray}
and the link fields must obey C-periodic boundary condition 
\cite{wiese}:
\begin{eqnarray}
U_{\mu} (s+N_{s} \hat{i}) = U^{\ast}_{\mu} (s) \hspace{3mm} 
{\rm ( complex \ conjugate )}.
\end{eqnarray}
The solid angle interpretation in the infinite-volume limit 
is applicable also in this boundary condition.
But the global Z(3) symmetry is broken explicitly in the case of SU(3).

\section{SU(2) case}

\begin{figure}
\vspace{-13mm}
\epsfxsize=0.40\textwidth
\begin{center}
\leavevmode
\epsfbox{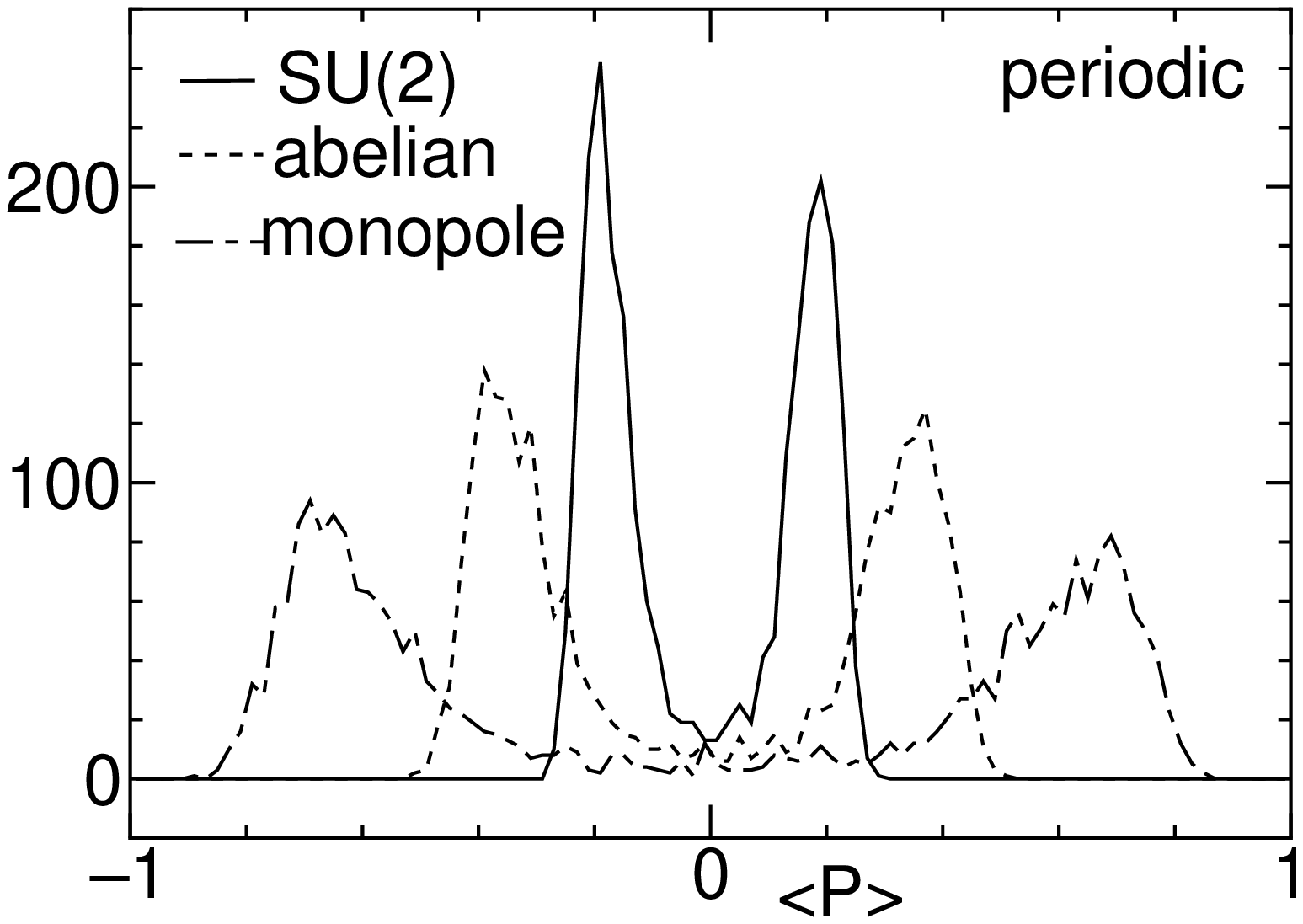}
\end{center}
\vspace{-35mm}
\epsfxsize=0.40\textwidth
\begin{center}
\leavevmode
\epsfbox{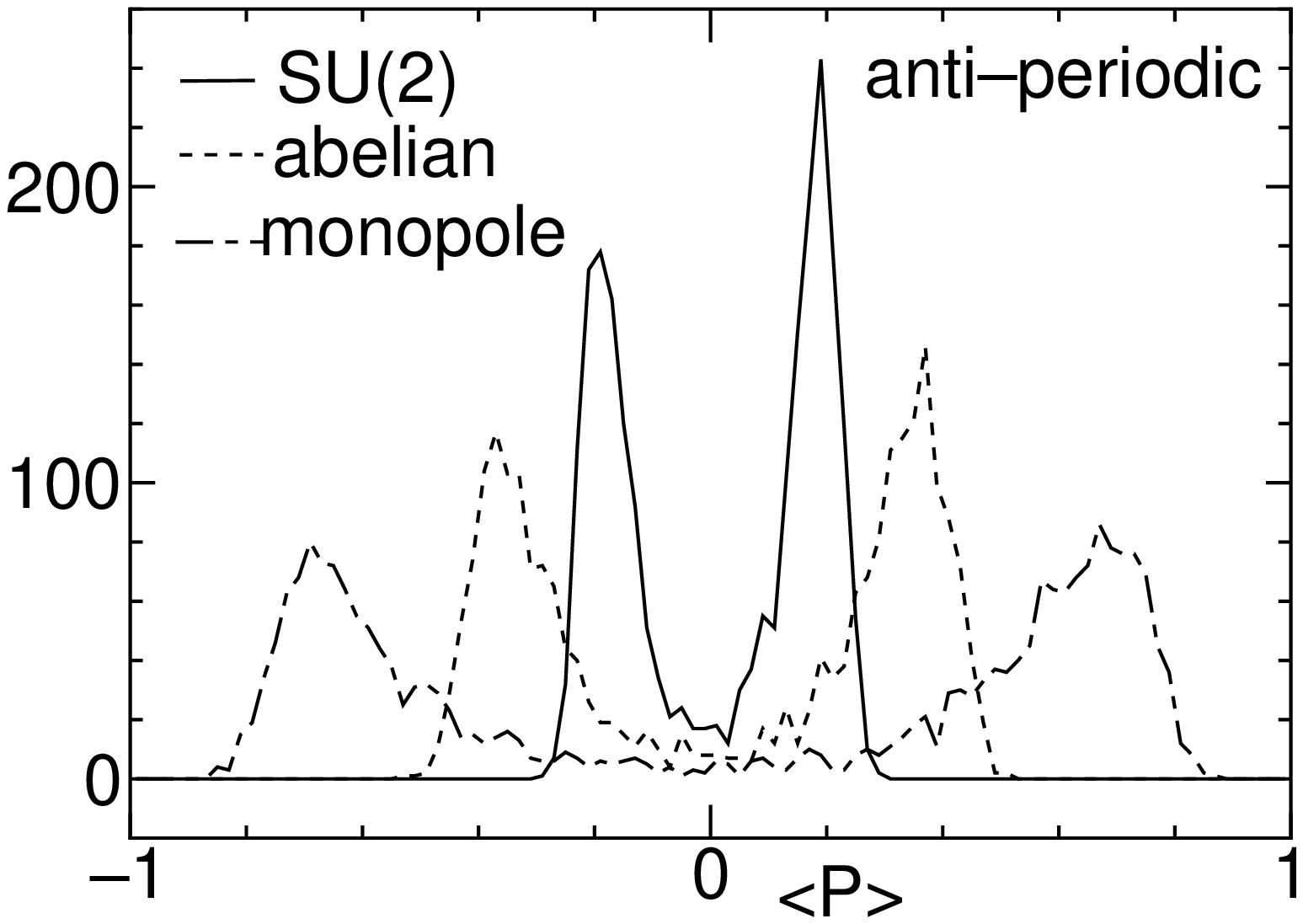}
\end{center}
\vspace{-35mm}
\caption{
The histogram of the Polyakov loop (averaged over $ \vec{x} $), 
the abelian Polyakov loop and the monopole part 
in the SU(2) gauge theory 
with periodic and anti-periodic boundary conditions.
($12^{3} \times 4$ lattice, $\beta = 2.32$) 
}
\vspace{-3mm}
\label{hist2}
\end{figure}

The histogram of the Polyakov loop (averaged over $ \vec{x} $) 
in the deconfinement phase of the SU(2) gauge theory 
is shown in Fig.\ref{hist2}.
The histogram of the Polyakov loop are similar in both boundary conditions, 
but the interpretation in terms of monopoles are quite different. 
In the case of the periodic boundary condition, 
the Z(2) symmetry is given by the contribution from the bubble of Dirac sheet 
due to the incompleteness of the Coulomb propagator. 
On the other hand, 
in the case of the anti-periodic (C-periodic) boundary condition,
there is a degree of freedom of introducing infinite size 
of Dirac sheet without changing the action.
The infinite size of Dirac sheet gives the Z(2) symmetry.
It gives $\Omega = 2 \pi$. If we add this Dirac string, 
$ \langle P_{m} \rangle \rightarrow - \langle P_{m} \rangle $. 
The action does not change by the transformation 
putting the infinite size of Dirac sheet on every time slice.

\section{SU(3) case}

\begin{figure}[tb]
\vspace{-10mm}
\hspace{-6mm}
\parbox{8cm}{
\epsfxsize=0.29\textwidth
\begin{flushleft}
\leavevmode
\epsfbox{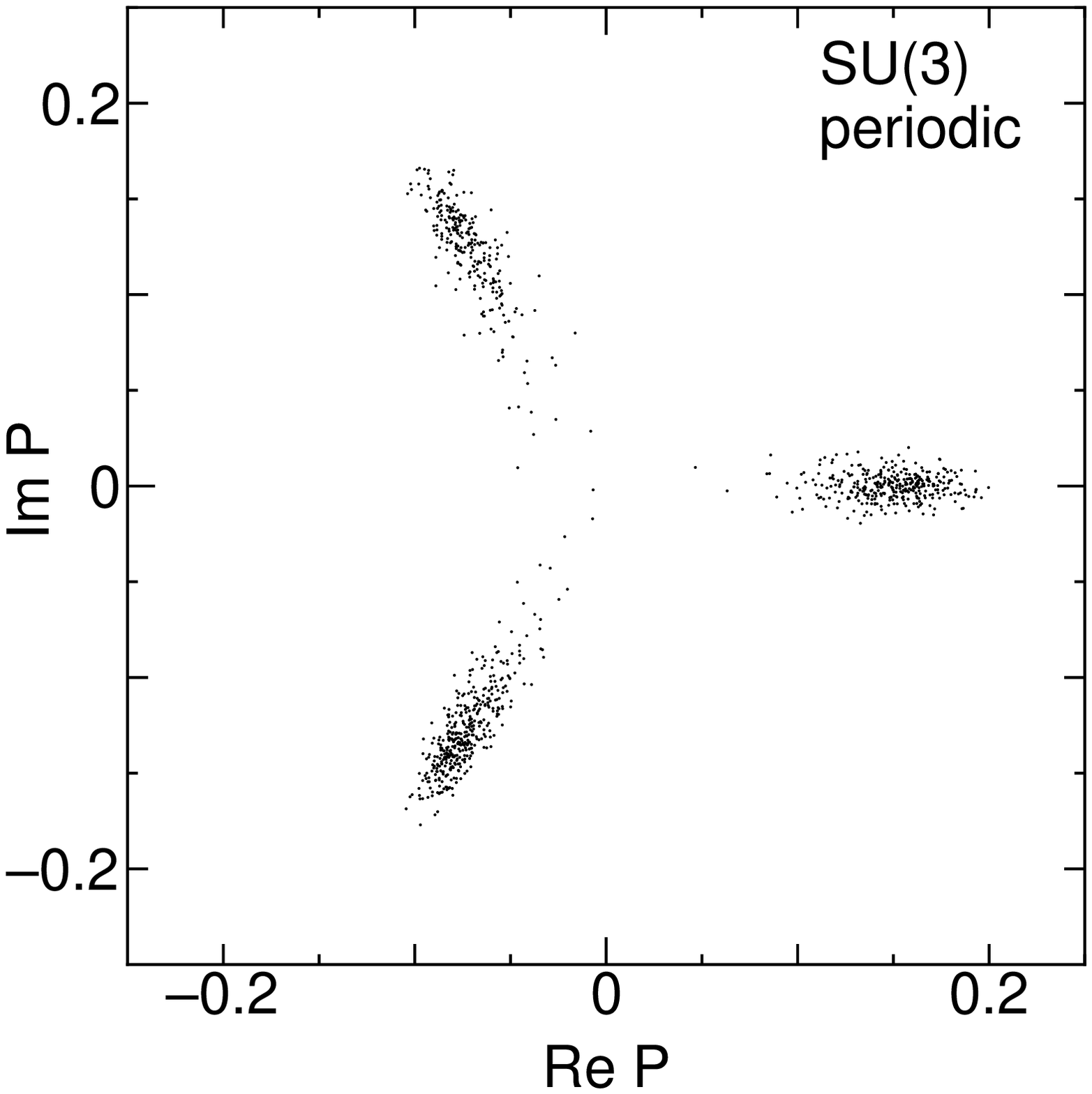}
\end{flushleft}
\vspace{-20mm}
\hspace{-8mm}
\epsfxsize=0.29\textwidth
\begin{flushleft}
\leavevmode
\epsfbox{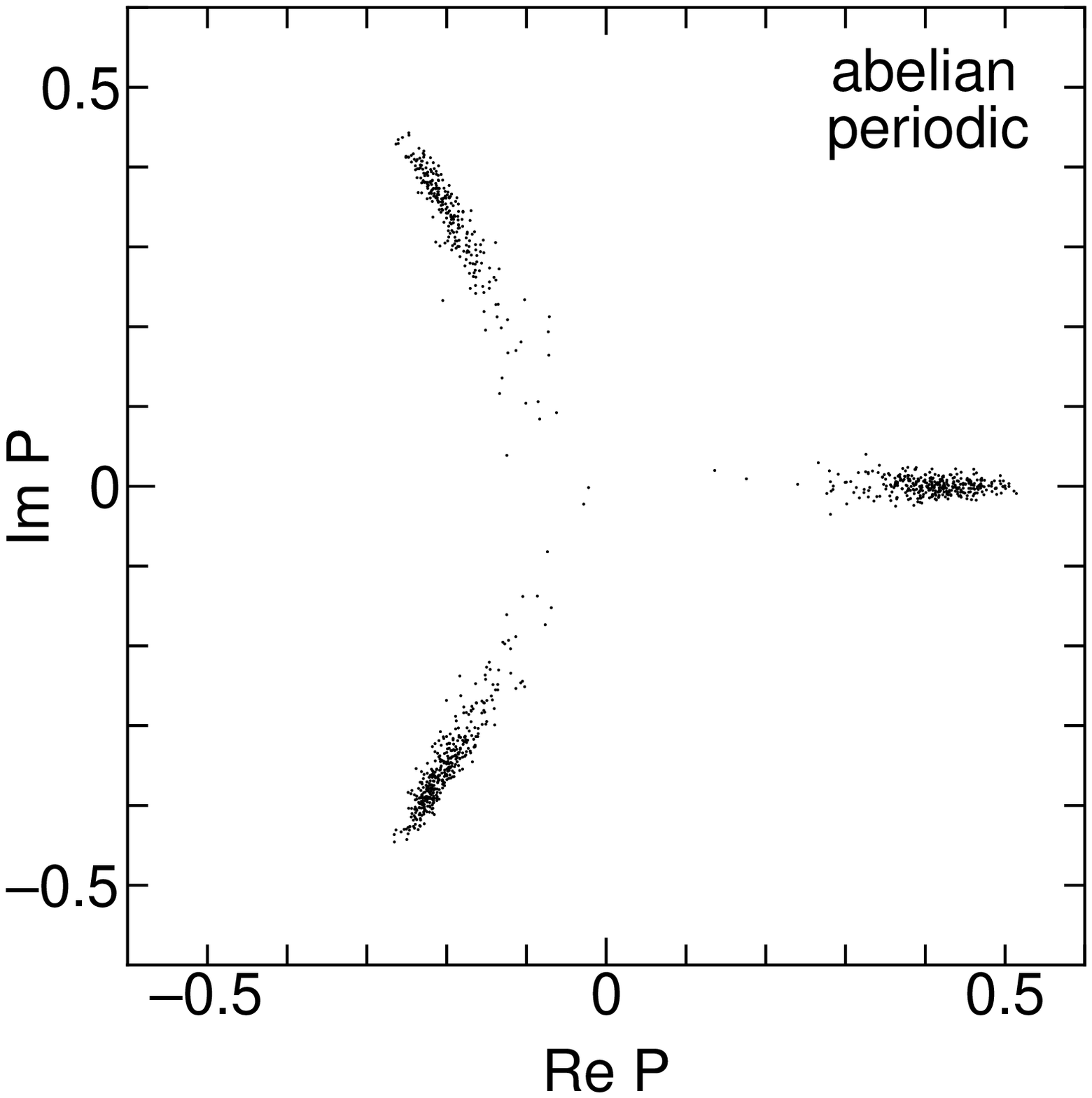}
\end{flushleft}
\vspace{-20mm}
\hspace{-8mm}
\epsfxsize=0.29\textwidth
\begin{flushleft}
\leavevmode
\epsfbox{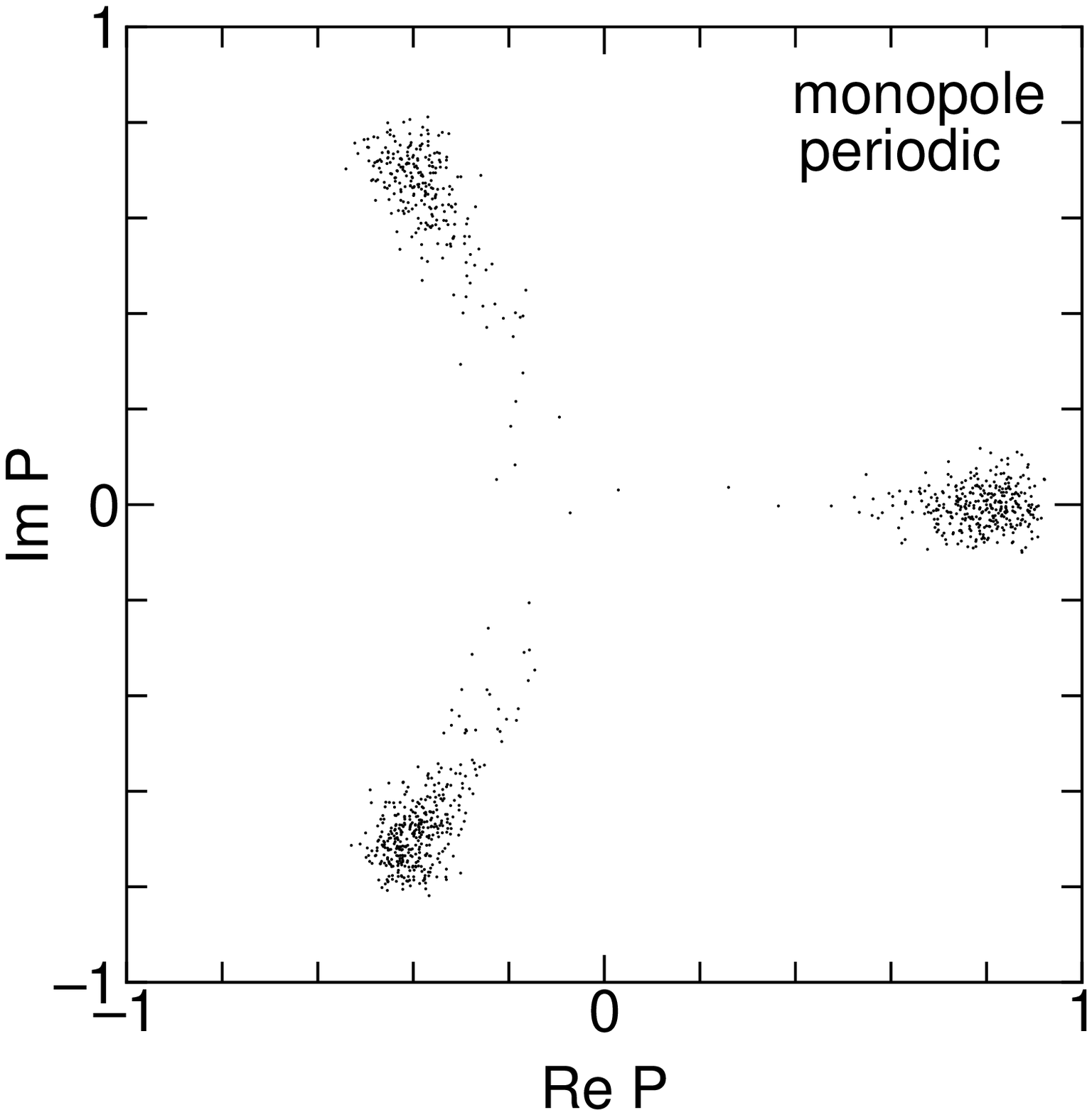}
\end{flushleft}
}
\vspace{-128mm}
\hspace{-3mm}
\parbox{8cm}{
\epsfxsize=0.29\textwidth
\begin{flushright}
\leavevmode
\epsfbox{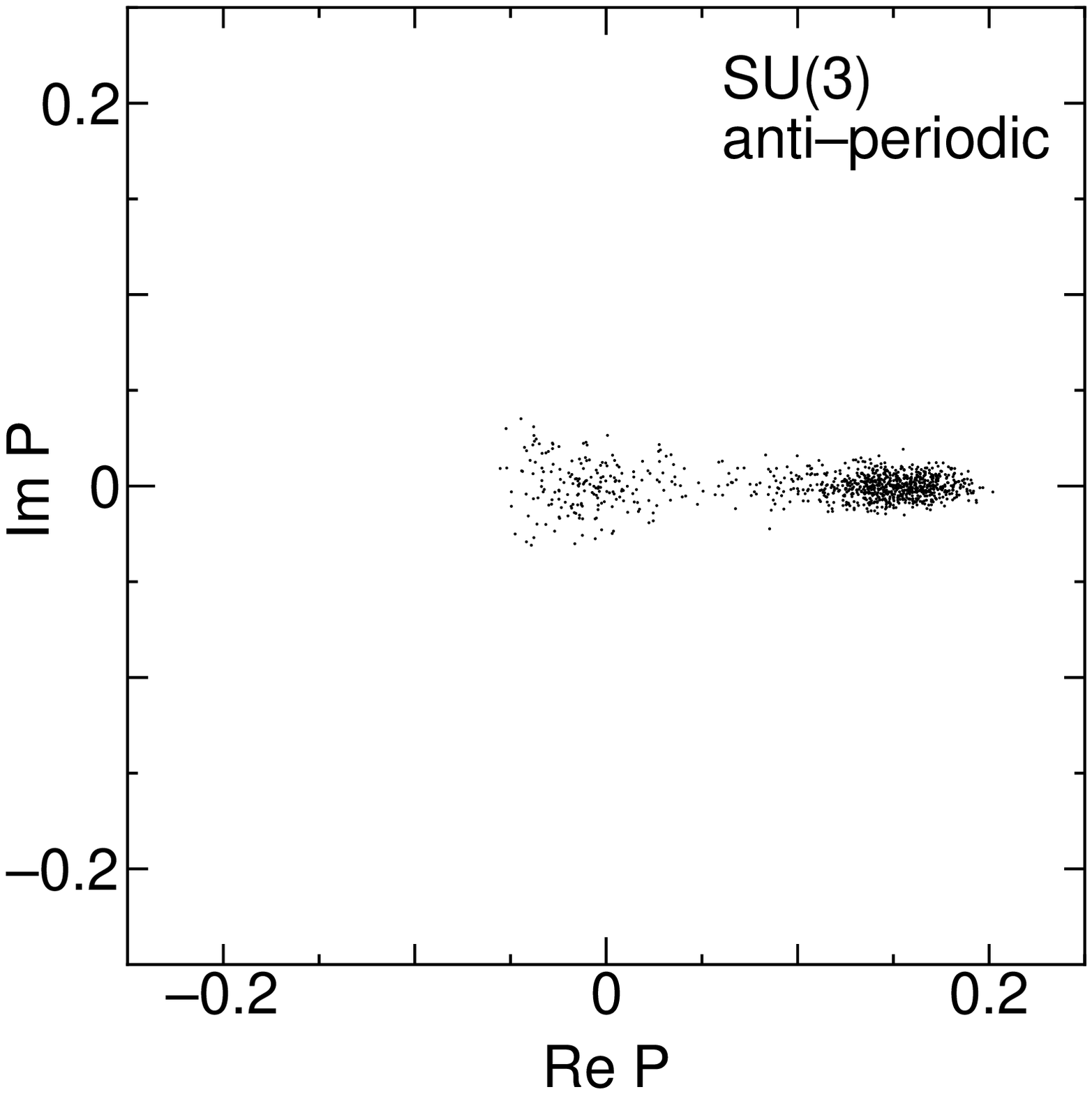}
\end{flushright}
\vspace{-20mm}
\hspace{-3mm}
\epsfxsize=0.29\textwidth
\begin{flushright}
\leavevmode
\epsfbox{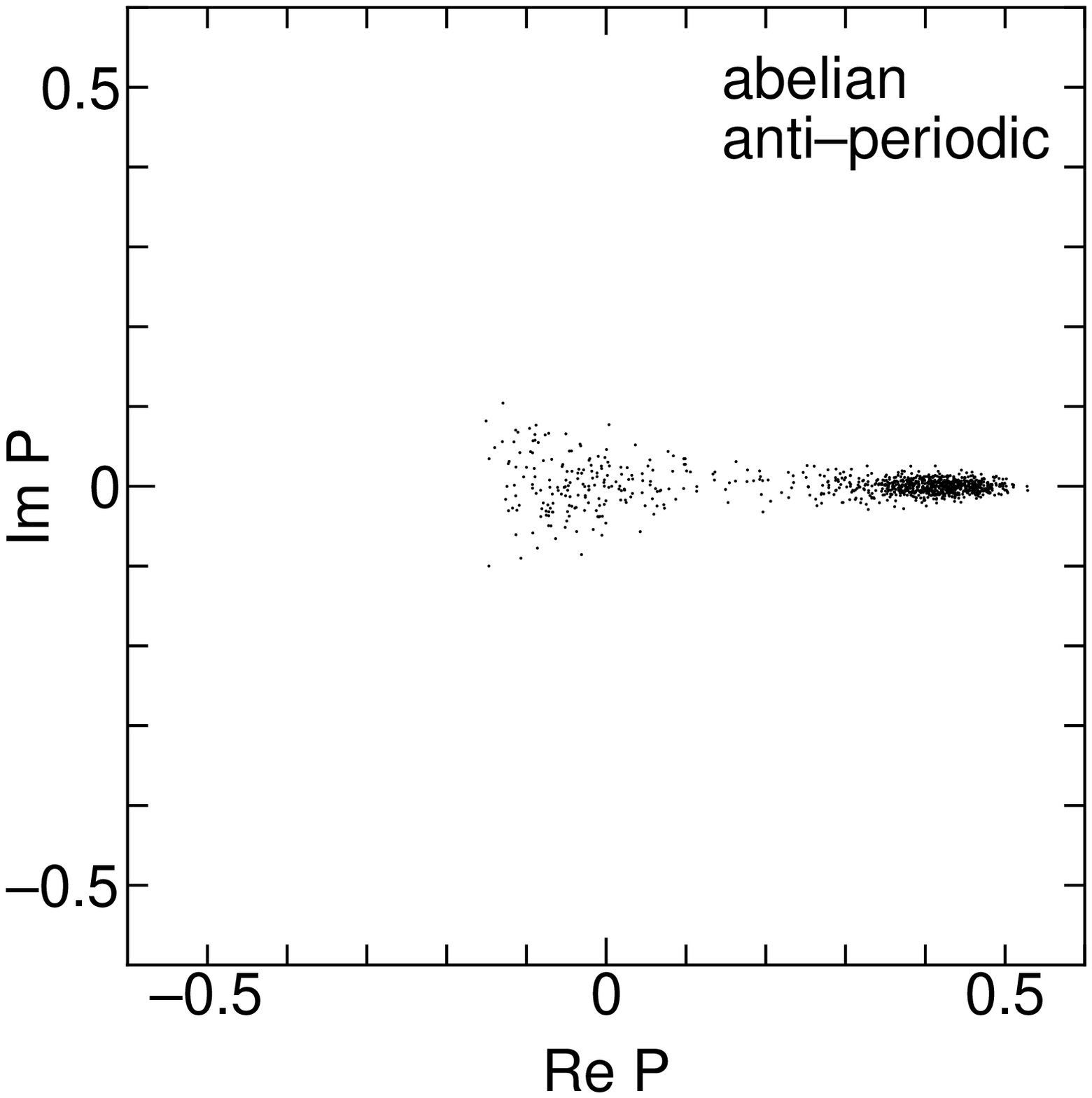}
\end{flushright}
\vspace{-20mm}
\hspace{-3mm}
\epsfxsize=0.29\textwidth
\begin{flushright}
\leavevmode
\epsfbox{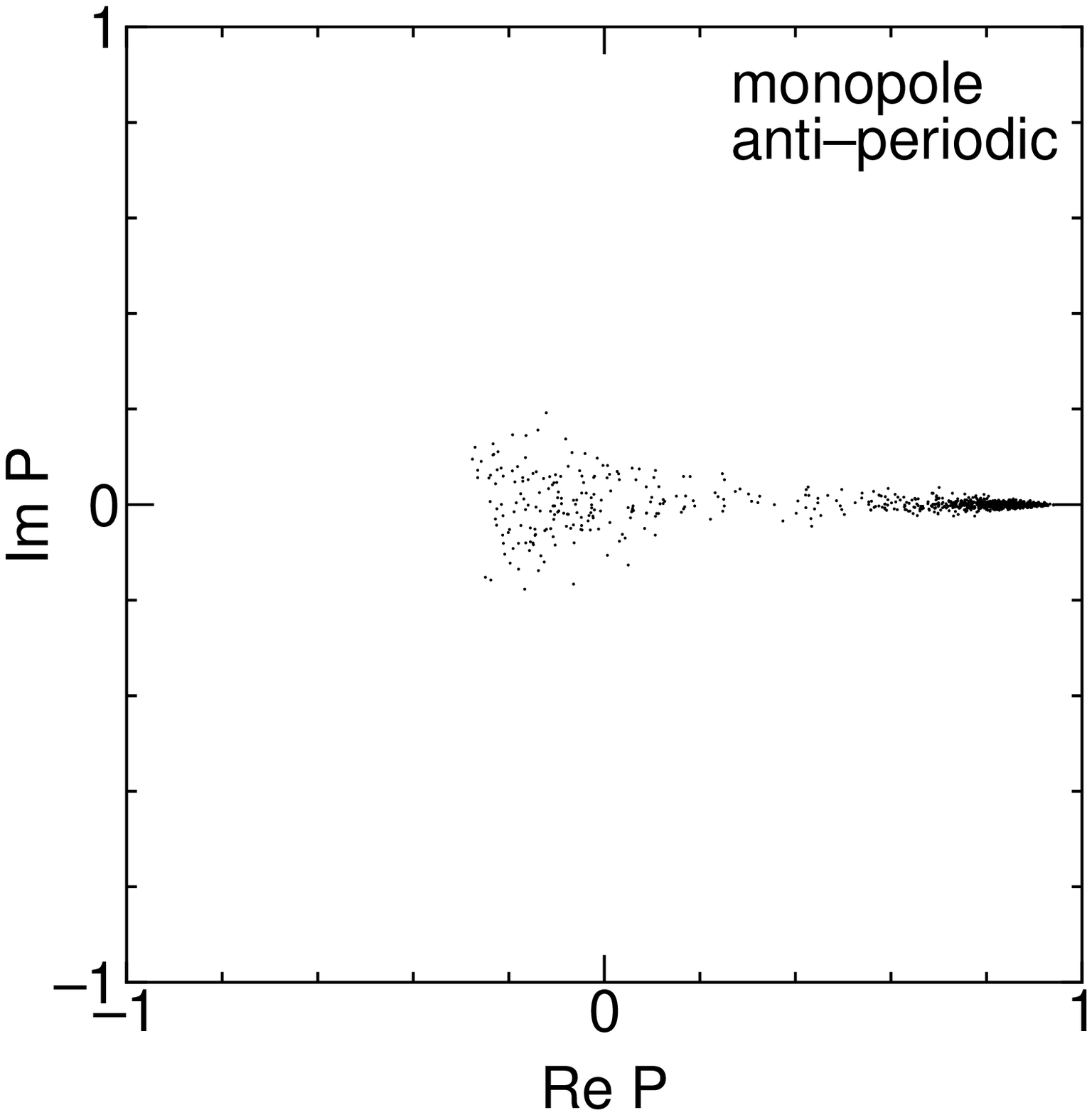}
\end{flushright}
}
\vspace{-17mm}
\caption{
The distribution of the Polyakov loop (averaged over $ \vec{x} $), 
the abelian Polyakov loop and the monopole part in the SU(3) gauge theory 
with periodic and anti-periodic boundary conditions. 
($12^{3} \times 4$ lattice, $\beta = 5.71$) 
}
\vspace{-3mm}
\label{plsu3}
\end{figure}

Fig. \ref{plsu3} is the distribution of the Polyakov loop 
(averaged over $ \vec{x} $) in the SU(3) gauge theory. 
In the periodic boundary condition, 
the global Z(3) symmetry is exact.
But the lattice Coulomb propagator can not be defined exactly. 
The effect of the bubble of Dirac sheet due to the incompleteness 
of the propagator gives the Z(3) phases as in SU(2) case.
The number of monopoles on each phase of the Polyakov loop 
is about the same in the deconfinement phase. 
On the other hand, in the anti-periodic (C-periodic) boundary condition, 
the solid angle explanation can be applicable, 
since the lattice Coulomb propagator is exact. 
The Z(3) symmetry is broken by the boundary condition. 
In the deconfinement phase, the Polyakov loop 
are distributed only in the vicinity of the real axis, 
This distribution can be understood by the 
monopole dynamics using the solid angle expression.

\vspace{5mm}
The author thanks T.Suzuki, Y.Matsubara, V.Bornyakov, S.Kitahara 
and F.Shoji for useful discussions and comments.

\end{document}